\documentclass[12pt]{article}
\usepackage{amssymb}
\usepackage[backend=biber]{biblatex}
\usepackage{hyperref}
\usepackage{graphicx}

\usepackage[utf8]{inputenc}
\DeclareUnicodeCharacter{2009}{~}
\bibliography{EDSBlois2017_Hanlon}

\newcommand{\xm}{$X_{\mathrm{max}}$}
\newcommand{\mxm}{$\langle X_{\mathrm{max}} \rangle$}

\newcommand{\spptotal}{$\sigma_{pp}^{\mathrm{tot}}$}
\newcommand{\spairinel}{$\sigma_{p-\mathrm{air}}^{\mathrm{inel}}$}

\def\Title#1{\begin{center} {\Large #1 } \end{center}}
\def\Author#1{\begin{center}{ \sc #1} \end{center}}
\def\Address#1{\begin{center}{ \it #1} \end{center}}
\def\andauth{\begin{center}{and} \end{center}}

\newenvironment{Abstract}{\begin{quotation}  }{\end{quotation}}
\newenvironment{Presented}{\begin{quotation} \begin{center} 
             PRESENTED AT\end{center}\bigskip 
      \begin{center}\begin{large}}{\end{large}\end{center} \end{quotation}}

\def\beq{\begin{equation}}
\def\eeq#1{\label{#1}\end{equation}}
\def\eeqn{\end{equation}}

\def\beqa{\begin{eqnarray}}
\def\eeqa#1{\label{#1}\end{eqnarray}}
\def\eeqan{\end{eqnarray}}

\let\bar=\overbar

\def\Dslash{\not{\hbox{\kern-4pt $D$}}}
\def\dslash{\not{\hbox{\kern-2pt $\del$}}}

\def\msb{{\bar{\ssstyle M \kern -1pt S}}}

\begin{document}
\begin{titlepage}

\vfill
\Title{Proton-Air Cross Section and Composition of Ultra High Energy
  Cosmic Rays Observed by Telescope Array}
\vfill
\Author{William Hanlon \andauth Rasha Abbasi \\
  \medskip\begin{center} on behalf of the Telescope Array Project\end{center}}
\Address{High Energy Astrophysics Institute \& Department of Physics
  and Astronomy, University of Utah, Salt Lake City, UT, USA}
\vfill
\begin{Abstract}
Ultra high energy cosmic rays (UHECRs) provide a natural source of
particles accelerated to energies beyond those that can be attained in
the laboratory. UHECRs have been observed with energies exceeding
$10^{20}$~eV, which is equivalent to 433~TeV in the center-of-momentum
frame. Using this natural source of particles physicists can extend
the measurement of the $pp$ cross section an order of magnitude above
what is achievable in the lab, possibly identifying hints of new
physics. The proton-air cross section and other properties of UHECR
QCD physics are also important in their own right to the study of the
sources and composition of UHECRs, but hadronic modelling at these
energies is still reliant upon phenomenological and the theoretical
extrapolations based upon terrestrial accelerator data. UHECR data can
be used to improve these extrapolations of the proton-air cross
section, but large uncertainties remain for other hadronic model
parameters. We present the most recent measurement of the inelastic
proton-air cross section at $\sqrt{s} = 95$~TeV measured by Telescope
Array using high quality \xm{} data collected in hybrid observing
mode. This measurement is also used to infer the total proton-proton
cross section.
\end{Abstract}
\vfill
\begin{Presented}
Presented at EDS Blois 2017, Prague, \\ Czech Republic, June 26-30, 2017
\end{Presented}
\vfill
\end{titlepage}
\def\thefootnote{\fnsymbol{footnote}}
\setcounter{footnote}{0}
%


\section{Introduction}
The total $pp$ cross section has been measured up to $\sqrt{s} =
8$~TeV by ATLAS~\cite{Aaboud:2016ijx} and
TOTEM~\cite{Antchev:2016vpy}. Nature provides us with cosmic
accelerators that inject particles with laboratory energies up to
$\gtrsim 10^{20}$~eV, equivalent to $\sqrt{s} \gtrsim 430$~TeV. Using
data from large cosmic ray experiments with exposures large enough to
collect sufficient statistics, we can measure \spptotal{} for energies
beyond what terrestrial accelerators can provide. The two largest
cosmic ray experiments currently operating have made such
measurements. The Pierre Auger Observatory has measured \spairinel{}
from which \spptotal{} can also be calculated and finds
$\sigma_{pp}^{\mathrm{tot}} = 133 \pm
13(\mathrm{stat})^{+17}_{-20}(\mathrm{sys}) \pm
16(\mathrm{Glauber})$~mb at $\sqrt{s} =
57$~TeV~\cite{Collaboration:2012wt}. Telescope Array has measured
\spairinel{} and \spptotal{} at $\sqrt{s} = 95$~TeV. This paper will
describe the method of measuring $pp$ cross section using cosmic rays,
the relation to air shower maximum size, and the results of Telescope
Arrays most recent measurements of \spairinel{} and
\spptotal. Understanding proton cross section is also important for
understanding the composition, and ultimately, the source of ultra
high energy cosmic rays (UHECRs). This is because \spairinel{} and the
depth of UHECR induced air shower maximum are related. At low energies
where the properties of \spairinel{} are constrained by accelerator
measurements, hadronic models used for UHECR simulations are tuned
using this information. The UHECR spectrum extends several orders of
magnitude beyond the highest energy attained by Earth bound
accelerators, so properties such as cross section, multiplicity, and
elasticity must be extrapolated over a large energy range. Studying
the energy dependence of these properties is important in constraining
these models at large energies, and reducing systematic uncertainties
in UHECR composition measurements.

\section{Experimental Method and Results}
\begin{figure}
  \centering
  \includegraphics[width=4in]{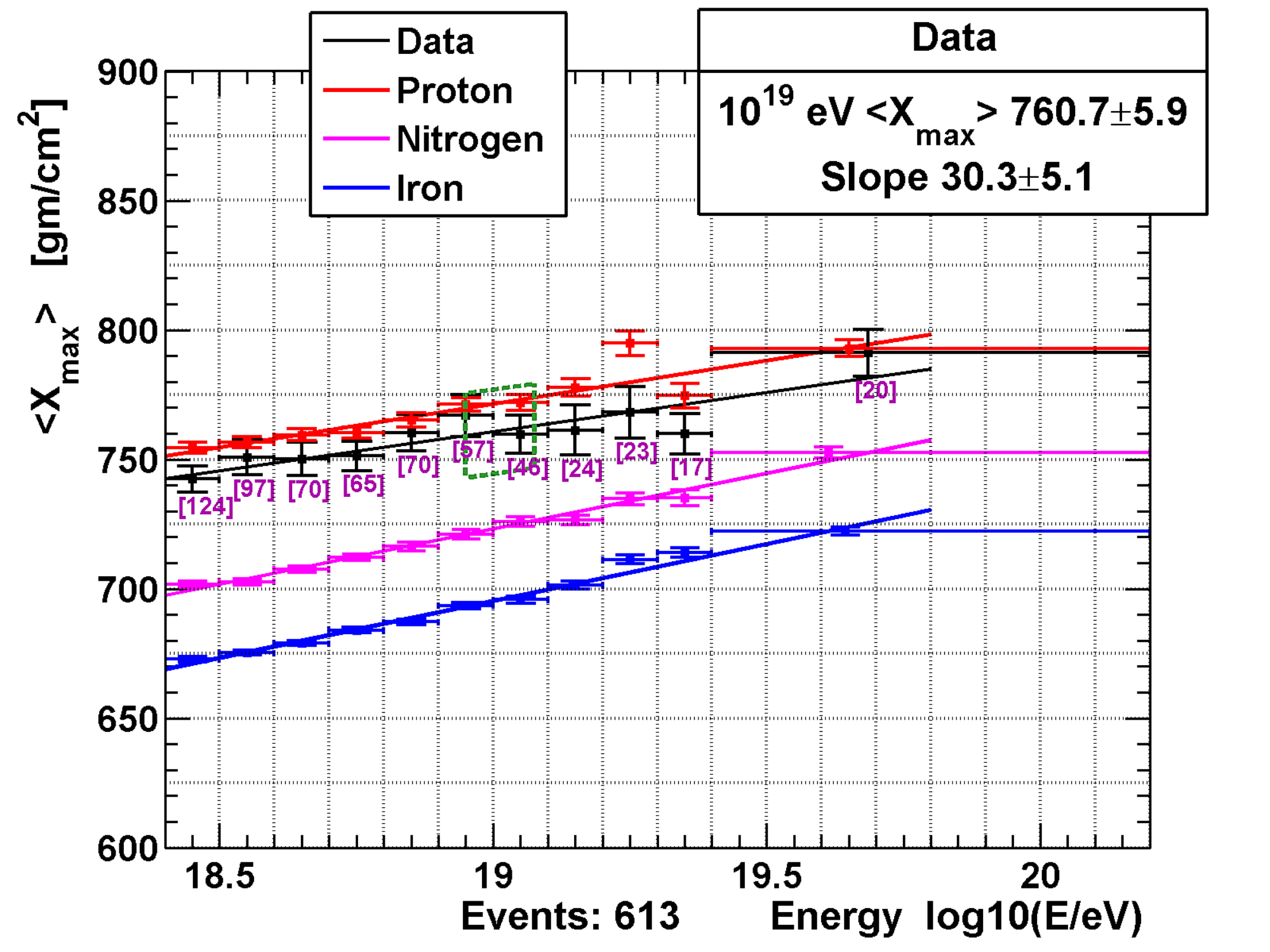}
  \caption{\mxm{} versus energy for five years of Telescope Array
    hybrid data observed using the Middle Drum fluorescence detector
    station. The data is compared to QGSJet~II-03 protons, nitrogen,
    and iron pure chemical compositions. The data used for TA's
    \spairinel{} and \spptotal{} measurements uses a subset of this data.}
  \label{fig:elongation}

\end{figure}

Telescope Array (TA) is a large cosmic ray observatory employing 507
scintillator surface detectors (SD) covering 700~km$^{2}$ and three
fluorescence detector (FD) stations overlooking the SD array. There
are a total of 48 FD telescopes that record the passage of UHECR
induced air showers as they pass through the atmosphere. This analysis
utilizes air shower data collected in hybrid mode, in which events
that simultaneously trigger both SDs and FDs are used to make precise
measurements of air shower maximum depth, also called \xm{} measured
in units of g/cm$^{2}$. A description of the TA experiment and its
equipment can be found in
\cite{AbuZayyad:2000uu,AbuZayyad:2012kk,Tameda:2009zza,Tokuno:2012mi}. The
hybrid data set used for this analysis contains 439 events from five
years of hybrid data recorded by the Middle Drum FD station collected
between May 2008 and May 2013~\cite{Abbasi:2014sfa}. The range of
(laboratory frame) energies accepted for this analysis is $10^{18.3}$
to $10^{19.3}$~eV. Figure~\ref{fig:elongation} shows the \mxm{} as a
function of energy measured by this analysis, as well as predictions
of \mxm{} expected for pure proton, nitrogen, and iron
compositions. The hadronic model used for the simulations is
QGSJet~II-03.

Hybrid reconstruction of UHECR induced air showers uses the
simultaneous measurement of geometry from the SD array, which
accurately measures the location of the shower core on the Earth's
surface, and the FDs, which measure the shower detector plane and
shower track vector in the atmosphere. Observation by these two
independent apparatuses provides excellent geometrical resolution,
which in turn provides good resolution in determining air shower
\xm. The \xm{} resolution for this analysis was $\sim 23$~g/cm$^{2}$.

\begin{figure}
  \centering
  \includegraphics[width=4in]{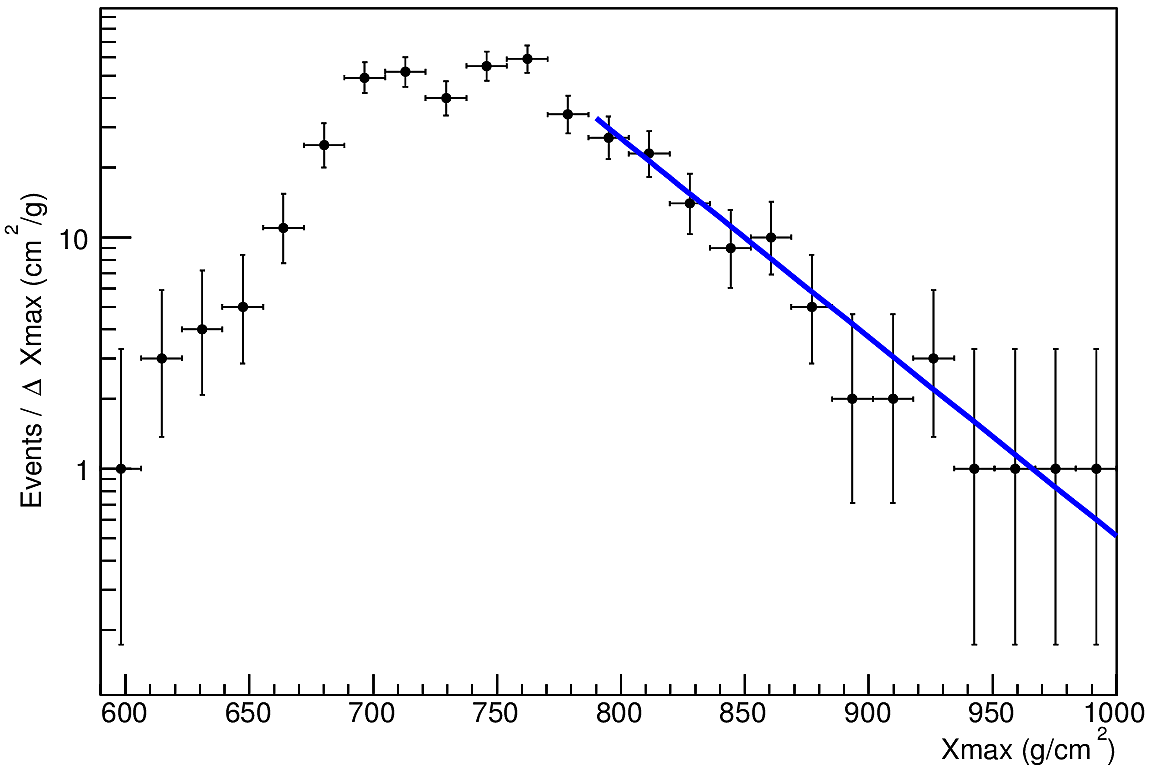}
  \caption{Number of data events per \xm{} bin used to measured
    \spairinel{}. The fit in the tail determines the slope which is
    related to the attenuation length of protons in air for $10^{18.3}
    < E < 10^{19.3}$~eV.}
  \label{fig:xmax_slope}
\end{figure}

Telescope Array measures \spairinel{} using the \textit{K-factor
  method}. This method relates the attenuation length of protons in
air to the exponential tail of the distribution of \xm{} observed from
many showers. Due to fluctuations in the first interaction of an UHECR
proton and an air molecule, the \xm{} distribution of protons exhibits
a long tail at deep \xm. This tail is fit to the functional form
$\exp(-X_{\mathrm{max}}/\Lambda_{\mathrm{m}})$, where
$\Lambda_{\mathrm{m}}$ is the attenuation
length. Figure~\ref{fig:xmax_slope} shows the \xm{} distribution of
the data and the fit to the tail found using this method.

$\Lambda_{\mathrm{m}}$ is related to \spairinel{} by the
relationship $\Lambda_{\mathrm{m}} = K\lambda_{p-\mathrm{air}} =
K\cdot(14.45m_{p}/\sigma_{p-\mathrm{air}}^{\mathrm{inel}})$. $K$ is
dependent upon the hadronic model chosen to simulate UHECR air
showers. To determine the value of $K$ used for this analysis,
simulated data sets of UHECR air showers reconstructed by TA were
used. Four hadronic models were used to generate the air showers:
QGSJet01~\cite{Kalmykov:1997te},
QGSJet~II-04~\cite{Ostapchenko:2010vb},
SIBYLL~2.1~\cite{Ahn:2009wx}, and EPOS~LHC~\cite{Pierog:2013ria}. The
values of $K$ and the \spairinel{} determined from the data are
\medskip
\begin{center}
      \begin{tabular}{l l l}
          Model         &$K$               &\spairinel{} (mb) \\
          \hline
          QGSJet01      &$1.22 \pm 0.01$   &$583.7 \pm 72.6$ \\
          QGSJet II-04  &$1.15 \pm 0.01$   &$550.3 \pm 68.5$ \\
          SIBYLL 2.1    &$1.18 \pm 0.01$   &$564.6 \pm 70.2$ \\
          EPOS LHC      &$1.19 \pm 0.01$   &$569.4 \pm 70.8$
      \end{tabular}
\end{center}
\medskip

Several sources of systematic uncertainty are measured. Model
dependence measured from the differences in \spairinel{} among the
four models used to determine $K$ and the proton-air inelastic cross
section contributes $\pm 17$~mb. The tail of the \xm{} distribution is
used to determine $K$ under the assumption that nearly all of those
events are initiated by a proton UHECR primary. The systematic effect
of contamination from several elements is examined with helium
producing the largest shift in \spairinel. Varying amounts of helium
contamination are examined as well as possible gamma ray
contamination. The systematic uncertainties studied and their values
are found to be

\medskip
\begin{center}
      \begin{tabular}{l l}
        Systematic source     & Systematic Uncertainty (mb) \\
          \hline
          Model dependence      &$\pm 17$ \\
          10\% helium           &$-9$ \\
          20\% helium           &$-18$ \\
          50\% helium           &$-42$ \\
          Gamma rays            &$+23$ \\
          \hline\hline
          Total (20\% helium)   &$(+29, -25)$
       \end{tabular}
\end{center}
\medskip

\begin{figure}
  \centering
  \includegraphics[width=4in]{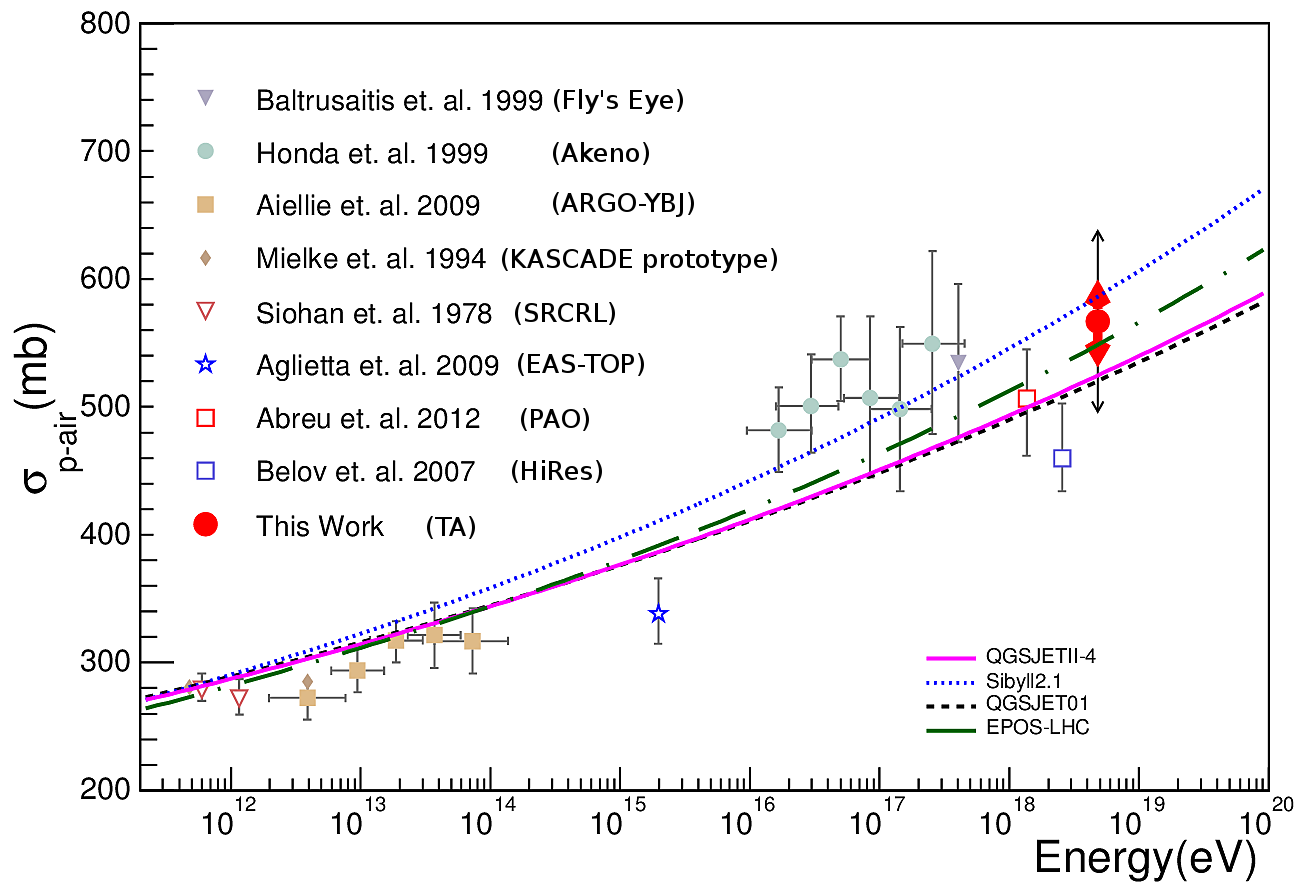}
  \caption{Telescope Array's measurement of $p$-air inelastic cross
    section. Statistical errors are indicated by the thin black lines
    around the data point, and systematic errors are indicated by the
    thick red lines. Other experimental results are shown, as well as
    predictions based on different hadronic models.}
  \label{fig:p_air_cross_section}
\end{figure}

A conservative estimate of up to 20\% helium contamination is used to
measure the total systematic uncertainty to be $(+29, -25)$~mb. The
final value of \spairinel{} is measured to be $567.0 \pm
70.5[\mathrm{stat}]^{+29}_{-25}[\mathrm{sys}]$~mb for $E =
10^{18.68}$, which is the mean energy of the data \xm{}
distribution. This corresponds to $\sqrt{s} =
95$~TeV. Figure~\ref{fig:p_air_cross_section} shows TA's result in
comparison to model predictions and other experimental measurements.

\begin{figure}
  \centering
  \includegraphics[width=4in]{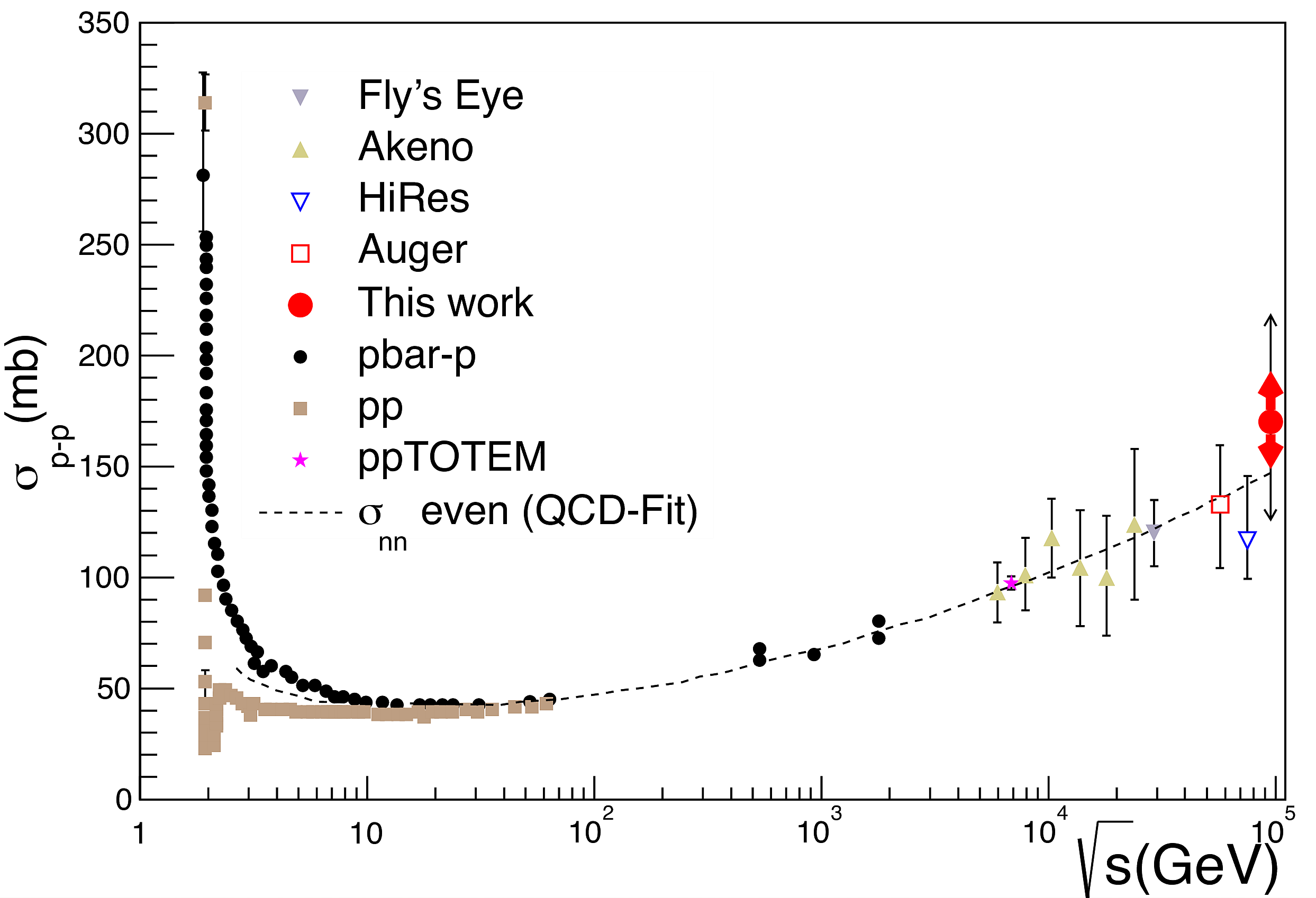}
  \caption{The total $pp$ cross section measured by Telescope Array at
    $\sqrt{s} = 95$~TeV. Statistical errors are indicated by the thin
    black lines around the data point, and systematic errors are indicated
    by the thick red lines. The dashed line shows the prediction of
    the QCD inspired fit by Block, Halzen, and Stanev.}
  \label{fig:p_p_cross_section}
\end{figure}

The total $pp$ cross section can be calculated using \spairinel{} and
Glauber formalism~\cite{Glauber:1970jm}. The relationship between
\spairinel{} and \spptotal{} is highly dependent on the forward
elastic scattering slope. We use the model developed by Block, Halzen,
and Stanev~\cite{Block:2011nr} which successfully describes Tevatron
data and is consistent with unitarity to calculate \spptotal{} from
our data. Propagating statistical and systematic errors from the
calculation of \spairinel{} we find $\sigma_{pp}^{\mathrm{tot}} =
170^{+48}_{-44}[\mathrm{stat}]^{+19}_{-17}[\mathrm{sys}]$~mb at
$\sqrt{s} = 95$~TeV. The placement of this result compared to other
measurements is shown in figure~\ref{fig:p_p_cross_section}. Further
details about this measurement are in \cite{Abbasi:2015fdr}.

\section{Conclusions}
Telescope Array has used five years of high quality \xm{} data to
measure the proton-air inelastic cross section and proton-proton total
cross section. Utilizing the distribution of \xm{} for events with
energies $10^{18.3} < E < 10^{19.3}$~eV and the $K$-factor method,
\spairinel{} is found to be $567.0 \pm
70.5[\mathrm{stat}]^{+29}_{-25}[\mathrm{sys}]$~mb at a mean lab energy
of $E = 10^{18.68}$~eV or $\sqrt{s} = 95$~TeV. This measurement can be
used to calculate the $pp$ total cross section using Glauber formalism
and the QCD inspired fit to Tevatron data. Using this method we
measure $\sigma_{pp}^{\mathrm{tot}} =
170^{+48}_{-44}[\mathrm{stat}]^{+19}_{-17}[\mathrm{sys}]$~mb at
$\sqrt{s} = 95$~TeV. Both results are consistent with model
predictions and other high energy measurements from other
experiments. Telescope array continues to collect \xm{} data and by
using other, larger hybrid data sets this result can be updated to
higher precision in the near future.

\printbibliography

\end{document}